\documentclass[11pt]{article}
\usepackage{amssymb,amsmath,amsfonts}
\usepackage[dvips]{graphicx}
\usepackage[dvips]{graphics}
\usepackage{eepic,epsfig}
\usepackage{verbatim}
\usepackage{color}
\usepackage{hyperref}
\usepackage[font=small,bf]{caption}
1.60cm \makeatletter \@addtoreset{equation}{section}

\makeatother

\begin{document}
\title{Non-Abelian cosmic strings in de Sitter and anti-de Sitter space}

\author{Ant\^onio de P\'adua Santos\thanks{E-mail:padua.santos@gmail.com} ,
 and Eug\^enio R. Bezerra de Mello\thanks {E-mail: emello@fisica.ufpb.br}\\
\\
\textit{Departamento de F\'{\i}sica, Universidade Federal da Para\'{\i}ba}\\
\textit{58.059-970, Caixa Postal 5.008, Jo\~{a}o Pessoa, PB, Brazil}\\
}

\maketitle
%
\begin{abstract}
In this paper we investigate the  non-Abelian cosmic string in de Sitter and anti-de Sitter spacetimes.
In order to do that we construct the complete  set of  equations of motion  considering the presence of a cosmological constant.
By using numerical analysis we provide the behavior of the Higgs and gauge fields
and also for the metric tensor for specific values of the physical parameters of  the theory.
For de Sitter case, we find the appearance of horizons that although being consequence of
the presence of the  cosmological constant  it strongly depends on the value of the  gravitational coupling.
In the  anti-de Sitter case, we find
that the system does not present  horizons. In fact  the new feature of this  system is
related with  the behavior of the $(00)$ and $(zz)$  components of the metric tensor. They present a
strongly increasing  for large distance from the string.

\end{abstract}
\bigskip


\bigskip
%
\section{Introduction}
\label{Int}
%

Our understanding about  the Universe  is based upon the standard cosmological model known as the Big Bang theory.
The main feature of the Big Bang Theory is the expansion of the Universe. Under this base, as the Universe expands it has been cooling.
During its cosmological expansion, the Universe underwent a series of phase transitions \cite{Kibble1976}.
These phase transitions are characterized by spontaneously broken gauge symmetries, and  have  important roles  in the
cosmological context \cite{Kibble1980}. They provide a mechanism for the formation of topological defects that can be described by
classical field theories whose configurations of vacuum have elegant and topologically stable solutions with relevant physical
implications. Such solutions are specified as domain wall, monopoles and cosmic strings among others \cite{Vilenkin-Shellard}.
Between them  cosmic strings are the most studied. They  can be considered as candidate to explain the temperature
anisotropies of cosmic microwave background (CMB) \cite{Ade:2013},  or associated to emission of gravitational wave and high-energy
cosmic rays \cite{Hindmarsh, Copeland}.

String-like solutions  were obtained  by Nielsen and  Olesen \cite{N-O} through a relativistic classical field theory
considering a system composed by Abelian and Non-Abelain gauge fields coupled with Higgs fields. In the system under consideration
a potential interaction which present non-trivial vacuum solution was taking into account. This potential is responsible for the
spontaneously broken of gauge symmetries. The authors were able to find static, cylindrically symmetric and 
stable solution from the equations of motion, which corresponds to a magnetic field along the $z-$direction. 
This solution was named {\bf vortex}.
Unfortunately the complete set of equations associated with this topological object is
non-linear and, in general, there is no closed solution for it. Only asymptotic expressions, for points near or
very far from the vortex's core, can be found for the Higgs and gauge fields. A more complex system is formed when
one decides to analyze the influence of this linear defect on the geometry of the spacetime.
This huge challenge was faced initially by Garfinkle \cite{DG} and two years later by Laguna and Matzner \cite{Laguna}
considering the Abelian version of the Nielsen and Olesen model. The authors have shown that there exists
a class of static cylindrically symmetric solutions of these equations representing a string;
moreover, they have showed that these solutions approach asymptotically to a Minkowski spacetime minus a wedge.
Linet in \cite{Linet87} analyzed a special kind of Abelian vortex solution that satisfy 
the BPS condition, and showed that for the case of infinity electric charge and Higgs field
self-coupling limit, it possible to obtain exact solutions for the metric tensor,
which is determined in terms of the linear energy density of the string.

The analysis of the spacetime geometry in the presence of an infinitely long, straight,
static, Abelian cosmic string formed during phase transitions at energy scales larger than the
grand-unified-theory scale were developed in \cite{Pablo,Ortiz}. For these supermassive configurations,
two different types of solutions were found:
one \cite{Pablo} in which the components of metric tensor $g_{tt}=g_{zz}$ vanish at finite distance from
the axis, and in the other where these components remain finite everywhere while $g_{\phi\phi}$ decreases
outside the core of the string. Although both types of geometries present different asymptotic behaviors, they are 
solutions of the same set of differential equations. This apparent contradiction was clarified in the 
papers \cite{Christensen,Brihaye-Lubo}, where the authors pointed out that the coexistence of two different kinds
of solutions are consequence of boundary conditions imposed on the metric fields. In fact the two different kinds
of asymptotic behaviors for the metric tensor correspond to the two different branches of cylindrically symmetric 
vacuum solutions of the Einstein equations \cite{Kramer}. The solution analyzed in \cite{Pablo} corresponds to 
the so called Melvin branch; and the case analyzed in \cite{Ortiz} corresponds to the so called string branch. The string branch solutions
are those of astrophysical interest, since they describe solutions with a deficit of planar angle \cite{Betti};
moreover, the Melvin branch has no flat spacetime counterpart. In \cite{Blanco} was discussesed how the presence of
multiple supermassive cosmic strings in the Abelian model can induce the spontaneous
compactication of the transverse space to a cosmic string, and to construct solutions where 
the gravitational background becomes regular everywhere.

In General Relativity, de Sitter (dS) and anti-de Sitter (AdS) spacetimes are  maximally symmetric solution
of Einstein's field equations in the presence of a positive and negative cosmological constant, $\Lambda$, respectively.
Due the symmetry of de Sitter and anti-de Sitter spacetimes,  numerous physical problems have been  exactly solved.
 In particular, astronomical observations of high redshift supernovae, galaxy clusters and cosmic microwave background
\cite{Riess, Perlmutter} indicate that at the present epoch we live in Universe that may be described by de Sitter spacetime. On the other
hand, anti-de Sitter spacetime plays an important role in theoretical physics such as the realization of the holographic principle
known as AdS/CFT correpondence \cite{Maldacena}. So, in the context of gravitating local cosmic string, a
natural question takes place: how does the presence of cosmological constant, positive or negative, modify the geometry of the spacetime produced by
an Abelian or non-Abelian cosmic string? The answer to this question is the main objective of the present analysis.

In fact the numerical analysis of the Abelian Nielsen and Olesen string minimally coupled to gravity including a positive cosmological constant have
been studied in \cite{eugenio2}. Moreover, the analysis of Abelian strings in a fixed background spacetime with positive cosmological constant 
has investigated in \cite{Ghezelbash,Yves}. In addition, the spherically symmetric topological defect named  global monopole \cite{BV}, was investigated in dS and AdS spacetimes by Li and Hao in \cite{Li} and by Bertrand at all in \cite{Bruno}.

In the paper by Nielsen and Olesen the non-Abelian string system is described by
a $SU(2)$ gauge invariant Lagrangian density composed by  gauge fields and two Higgs sectors. A potential responsible for the
spontaneously gauge symmetry broken is present. The analysis of the  non-Abelian Nielsen and Olesen string 
and its influence on the geometry of the spacetime, was only recently considered in \cite{Ant_Eugenio}.
In this analysis it was not take into account the presence of
any cosmological constant. All the modifications in the Minkowski spacetime were caused by the defect.  
So, as an addition motivation to develop this work, we would like to complete this analysis considering now 
that the non-Abelian string, and also tne Abelian one, are in dS and AdS spacetimes. 

This  paper is organized as follows: In section \ref{Model} we  present the non-Abelian Higgs model in de Sitter and anti-de Sitter
spaces and analyze the conditions that the physical parameters contained in the potential should satisfy in order the
system presents stable topological solutions. Also we present the ansatz for the Higgs and gauge fields, and for the
metric tensor.  The equations of motion and boundary
conditions are presented in section \ref{Motion}. In section \ref{Numerical} we provide
our numerical results, exhibiting the behaviors of the Higgs, gauge and metric fields as
functions of the distance to the core of the string. Moreover, we present a comparison of the non-Abelian system 
in Minkowski, de Sitter and anti-de Sitter spaces, and pointed out the most relevant aspects 
that distinguish the behaviors of those fields in the presence/absence of cosmological 
constant. Finally in section \ref{Concl} we give our conclusions.
%
%
\section{The Model}
\label{Model}
%
In previous work \cite{Ant_Eugenio} it has been studied the behavior of gravitating non-Abelian strings in the absence of cosmological constant.
 In that paper it was mainly considered the planar angle deficit in the spacetime caused by the string
  and the  the energy density by unit length associated with this
  system,  and compare both quantity, separately, with the corresponding ones for the Abelian string.
The aim of this paper is to examine the influence of the cosmological constant in the non-Abelian
and Abelian cosmic strings spacetimes. For the present purposes, we introduce the cosmological constant in the model
described by the following action, $S$:
\begin{equation}
 S = \int d^4x \sqrt{-g}\left(\frac{1}{16\pi G}(R - 2\Lambda) + \mathcal{L}_m\right), \label{eqAction}
\end{equation}
 where $R$ is the Ricci scalar, $G$ denotes the Newton's constant and
 $\Lambda$ is the cosmological constant\footnote{For de Sitter space  
 $\Lambda > 0$ and for anti-de Sitter space $\Lambda < 0$.}.
 The  matter Lagrangian density of the non-Abelian Higgs model is given by
 \begin{equation}
 \mathcal{L}_m = -\frac{1}{4}F^a_{\mu \nu}F^{\mu \nu a} + \frac{1}{2}(D_{\mu}\varphi^a)^2 +
 \frac{1}{2}(D_{\mu}\chi^a)^2 - V(\varphi^a, \chi^a),
 \quad a = 1, 2, 3; \label{eqLagrangian}
\end{equation}
where $F^{a}_{\mu \nu}$ denotes the field strength tensor,
\begin{equation}
 F^{a}_{\mu \nu} = \partial_{\mu} A^a_{\nu} - \partial_{\nu} A^a_{\mu} + e \epsilon^{abc}A^b_{\mu}A^c_{\nu}.
\end{equation}
The covariant derivative is given by $D_{\mu}\varphi^a = \partial_{\mu} \varphi^a + e \epsilon^{abc}A^b_{\mu}\varphi^c$, where the
latin indices denote the internal gauge groups. $A^b_{\mu}$ is the $SU(2)$ gauge potential and $e$
the gauge coupling constant.
The interaction potential, $V(\varphi^a, \chi^a)$, is defined by the expression
\begin{eqnarray}
V(\varphi^a, \chi^a) &  =  &\frac{\lambda_1}{4}\left[(\varphi^a)^2 - \eta_1^2\right]^2
+ \frac{\lambda_2}{4}\left[(\chi^a)^2- \eta_2^2\right]^2 \nonumber \\
& & + \frac{\lambda_3}{2}\left[(\varphi^a)^2 - \eta_1^2\right]\left[(\chi^a)^2 - \eta_2^2\right],
\end{eqnarray}
where the $\lambda_1 $ and $\lambda_2 $ are the Higgs fields self-coupling positive constants
and $\lambda_3 $ is the coupling constant between both  bosonic
sectors. $\eta_1 $ and $\eta_2$ are parameters corresponding to the energies scales where the gauge
symmetry is broken. The potential above has different
properties according to the sign of $\Delta \equiv \lambda_1 \lambda_2 - \lambda_3^2$  \cite{Ant_Eugenio}:
\begin{itemize}
 \item For $\Delta>0 $, the potential has positive value and its minimum is attained
 for $(\varphi^a)^2 = \eta_1^2 $ and $(\chi^a)^2 = \eta_2^2$.
  \item For $\Delta<0$, these configuration lead to saddle points and two minima occur for:
\begin{equation}
 (\varphi^a)^2 = 0, \quad (\chi^a)^2 = \eta_2^2 + \frac{\lambda_3}{\lambda_2}\eta_1^2
\end{equation}
and
\begin{equation}
 (\chi^a)^2 = 0, \quad (\varphi^a)^2 = \eta_1^2 + \frac{\lambda_3}{\lambda_1}\eta_2^2.
\end{equation}
The values of the potential for these cases are, respectively,
\begin{equation}
V_{min}=\frac{\eta_1^4}{4\lambda_2}\Delta \quad \textrm{and} \quad V_{min}=\frac{\eta_2^4}{4\lambda_1}\Delta.
\end{equation}
Both values for $V_{min} $ are negatives, since $\Delta<0$.
\end{itemize}

%
\subsection{The Ansatz}
\label{Ansatz}

First let us consider the most general,
cylindrically symmetric line element invariant under boosts along z-direction.
By using cylindrical coordinates, this line element is given by:
\begin{equation}
 ds^2 = N^2(\rho)dt^2 - d\rho^2 - L^2(\rho)d\phi^2 - N^2(\rho)dz^2 \  .
 \label{ds}
\end{equation}
For this metric, the non-vanishing components of the Ricci tensor, $R_{\mu\nu}$,  are:
\begin{equation}
 R_{tt} = - R_{zz} = \frac{NLN''+ NN'L' + L(N')^2}{L} \  ,
\end{equation}
\begin{equation}
 R_{\rho\rho} = \frac{2LN'' + NL''}{NL} \  ,
\end{equation}
\begin{equation}
 R_{\phi \phi} = \frac{L(2N'L' + NL'')}{N}   ,
\end{equation}
where the primes denotes derivative with respect to $\rho$.

For the  Higgs and gauge fields we have the following expressions \cite{Vega}:
\begin{equation}
 \varphi^a(\rho) = f(\rho)
\left( \begin{array}{c}
 \cos(\phi)\\
 \sin(\phi)\\
   0
\end{array} \right) \  ,
\end{equation}
\begin{equation}
 \chi^a(\rho) = g(\rho)
\left( \begin{array}{c}
 -\sin(\phi)\\
 \cos(\phi)\\
   0
\end{array} \right)  \  ,
\end{equation}
\begin{equation}
\vec{A}^a(\rho) = \hat{\phi}\left(\frac{1-H(\rho)}{e\rho}\right)\delta_{a,3}
\end{equation}
and
\begin{equation}
 {A}^a_t(\rho) = 0 \  , \quad  a = 1, 2, 3 \  .
\end{equation}
From the above expressions we can see that both iso-vector bosonic fields satisfy the orthogonality condition,  $\varphi^a\chi^a=0$.
%
\section{Equation of Motion}
\label{Motion}
%
In this paper we shall use the same notation as in \cite{Ant_Eugenio} for the dimensionless variables and functions, as shown below:
\begin{equation}
 x = \sqrt{\lambda_1}\eta_1\rho, \quad f(\rho) = \eta_1X(x),\ g(\rho) = \eta_1Y(x),
  \ \  L(x) = \sqrt{\lambda_1}\eta_1 L(\rho).
\end{equation}
Adopting these notations the Lagrangian density will depend only on  dimensionless variables and parameters:
\begin{equation}
 \alpha = \frac{e^2}{\lambda_1}, \quad q = \frac{\eta_1}{\eta_2}, \quad \beta^2_i =
 \frac{\lambda_i}{\lambda_1}, \quad i = 1, 2, 3 \  ,  \
\gamma = \kappa \eta_1^2,\ \bar{\Lambda}=\frac{\Lambda}{\eta_1^2\lambda_1} \ {\rm and} \  \kappa = 8\pi G \ .
\end{equation}
For the de Sitter or anti-de Sitter spacetime, it is convenient to use the Einstein field equations in the form
\begin{equation}
 R_{\mu\nu} = -\kappa\left(T_{\mu\nu} -\frac{1}{2}g_{\mu\nu}T\right) + {\Lambda}g_{\mu\nu},
 \quad \textrm{with} \quad T = g^{\mu\nu}T_{\mu\nu}  \quad
\textrm{and}\quad \mu, \nu = t, x, \phi, z \  .  \label{eqEinstein}
\end{equation}
For the energy-momentum tensor associated with the mater field we use the usual definition given below,
\begin{equation}
T_{\mu\nu} = \frac{2}{\sqrt{-g}}\frac{\delta S}{\delta g^{\mu\nu}}, \quad g = \textrm{det}(g_{\mu\nu}).
\end{equation}

Varying the action (\ref{eqAction}) with respect to matter fields, we obtain the Euler-Lagrange equations below,
\begin{equation}
 \frac{(N^2LX')'}{N^2L}= X\Biggl[X^2-1 + \beta_3^2\left(Y^2-q^2\right) + \frac{H^2}{L^2}\Biggr] \  , \label{eq1}
\end{equation}
\begin{equation}
 \frac{(N^2LY')'}{N^2L}= Y\Biggl[\beta_3^2\left(X^2-1\right) + \beta_2^2\left(Y^2-q^2\right) + \frac{H^2}{L^2}\Biggr], \label{eq2}
\end{equation}
\begin{equation}
 \frac{L}{N^2} \left(\frac{N^2H'}{L}\right)' = \alpha \bigl(X^2 + Y^2)H  \  .  \label{eq3}
\end{equation}
As to the Einstein equations (\ref{eqEinstein}), we obtain:
\begin{equation}
 \frac{\left(LNN'\right)'}{N^2L} = -\bar{\Lambda}+\gamma \Biggl[\frac{H'^2}{2\alpha L^2} -\frac{1}{4}\left(X^2-1\right)^2 -\frac{\beta_2^2}{4}\left(Y^2-q^2\right)^2-
 \frac{\beta_3^2}{2}\left(X^2 -1\right)\left(Y^2-q^2\right)\Biggr]  \label{eq4}
\end{equation}
and
\begin{eqnarray}
\frac{\left(N^2L'\right)'}{N^2L}&  = & -\bar{\Lambda} -\gamma \Biggl[\frac{H'^2}{2\alpha L^2} + \left(X^2
+Y^2\right)\frac{H^2}{L^2}+\frac{1}{4}\left(X^2-1\right)^2
 +\frac{\beta_2^2}{4}\left(Y^2-q^2\right)^2 \nonumber \\
 & & + \frac{\beta_3^2}{2}\left(X^2 -1\right)\left(Y^2-q^2\right) \Biggr]  \  .  \label{eq5}
\end{eqnarray}
The primes in the equations (\ref{eq1}) - (\ref{eq5}) stand for derivatives with respect to $x$.
As we can see this set of non-linear coupled differential equations is a hard sistem to be analyzed. We shall
leave this task for the next section.
Defining $u = \sqrt{-g} = N^2L$, we obtain the following equation:

\begin{eqnarray}
\frac{u''(x)}{u(x)} & = & -3\bar{\Lambda}-\gamma\Biggl[-\frac{H'^2}{2\alpha L^2} + \left(X^2
+Y^2\right)\frac{H^2}{L^2}+\frac{3}{4}\left(X^2-1\right)^2
 +\frac{3\beta_2^2}{4}\left(Y^2-q^2\right)^2 \nonumber \\
& &  + \frac{3\beta_3^2}{2}\left(X^2 -1\right)\left(Y^2-q^2\right) \Bigg]. \label{u}
\end{eqnarray}

Before to finish this subsection, we would like to point out  that the set of differential equation above,
reduces itself to the corresponding one for the Abelian Higgs model by taking $\beta_2=\beta_3=0$ and setting  one of the
Higgs field equal to zero. Because one of our objective is to compare the non-Abelian results with the corresponding one for
the Abelian case,  we shall take, when necessary, the bosonic field $\chi=0$, which, in terms of dimensionless functions,
corresponds to take $Y=0$.

%
\subsection{Boundary conditions}
%

The boundary conditions imposed on the fields at origin are determined by the requirements of regularity at this point. However,
the sign of the cosmological constant, $\bar\Lambda$, will establish different kinds of the boundary conditions for the matter
and gauge fields at large distance.
\begin{itemize}
\item  For de Sitter space $(\bar\Lambda > 0)$, the boudary conditions for the matter and gauge fields are:
\begin{equation}
     H(0) = 1, \ X(0) = 0, \ Y(0) = 0. \label{eqbound1}
\end{equation}
As we shall see,  the cosmological constant will provide a cosmological  horizon for the metric tensor.
Then we must integrate the equations until this value of the coordinate,
$x=x_0$, in order to have the core of the cosmic string located inside the horizon. So, we require:
\begin{equation}
     X(x=x_0) = 1, \ Y(x=x_0) = \frac{\eta_2}{\eta_1} = q,  \  H(x=x_0) = 0.   \label{eqbound2}
\end{equation}

\item  For anti-de Sitter space $(\bar\Lambda < 0)$, the cosmological horizon does not appear.  Therefore the boundary conditions
for the matter and gauge fields are:

\begin{equation}
     H(0) = 1; \quad H(\infty) = 0 \  , \label{eqbound3}
\end{equation}
\begin{equation}
     X(0) = 0, \quad X(\infty) = 1,  \quad  Y(0) = 0,  \quad Y(\infty) = \frac{\eta_2}{\eta_1} = q. \label{eqbound4}
\end{equation}
\end{itemize}
The boundary conditions for the metric fields are
\begin{equation}
     N(0) = 1, \quad N'(0) = 0 \  , \quad L(0) = 0, \quad L'(0) = 1  \label{eqbound5}
\end{equation}
in both spaces.

%
\subsection{Vacuum solution}
\label{vaccum}
%
The vacuum solution of our system is attained by setting $X(x) =1, \  Y(x) = q$ and $ H(x) = 0$ into the Eq. \eqref{u}. So, we have:
\begin{itemize}
\item For de Sitter spacetime ($\bar\Lambda > 0$), we get

\begin{equation}
 N^2(x)L(x) = A_1\sin(\sqrt{3\bar\Lambda}x) + B_1\cos(\sqrt{3\bar\Lambda}x) \ .
\end{equation}
\vspace{0.5cm}
Using the boundary condition Eq. (\ref{eqbound5}), we find the following solution

\begin{equation}
N^2(x)L(x) =\frac{1}{\sqrt{3\bar\Lambda}} \sin(\sqrt{3\bar\Lambda} x).
\end{equation}
Following the method sugested by Linet \cite{Linet}, we find the solutions
\begin{equation}
\label{NdS}
 N(x) = \cos^{2/3}\Biggl(\sqrt{3\bar\Lambda}\frac{x}{2}\Biggr)
\end{equation}
and
\begin{equation}
\label{LdS}
 L(x) = \frac{2^{2/3}}{\sqrt{3\bar\Lambda}}\Biggl[\sin(\sqrt{3\bar\Lambda}x)\Biggr]^{1/3}
 \Biggl[\tan\biggl(\sqrt{3\bar\Lambda}\frac{x}{2}\biggr)\Biggr]^{2/3}.
\end{equation}
\item For anti-de Sitter spacetime ($\bar\Lambda < 0$), we get

\begin{equation}
 N^2(x)L(x) = A_2\exp(\sqrt{3|\bar\Lambda|}x) + B_2\exp(-\sqrt{3|\bar\Lambda|}x)
\end{equation}
\vspace{0.5cm}
Using the boundary condition Eq. (\ref{eqbound5}), we find the following solution
\begin{equation}
N^2(x)L(x) =\frac{1}{\sqrt{3|\bar\Lambda|}} \sinh(\sqrt{3|\bar\Lambda|} x).
\end{equation}
By means of \cite{Linet}, we find the solutions
\begin{equation}
\label{NAdS}
 N(x) = \cosh^{2/3}\Biggl(\sqrt{3|\bar\Lambda|}\frac{x}{2}\Biggr)
\end{equation}
and
\begin{equation}
\label{LAdS}
 L(x) = \frac{2^{2/3}}{\sqrt{3|\bar\Lambda|}}\Biggl[\sinh(\sqrt{3|\bar\Lambda|}x)\Biggr]^{1/3}
 \Biggl[\tanh\biggl(\sqrt{3|\bar\Lambda|}\frac{x}{2}\biggr)\Biggr]^{2/3}.
\end{equation}
\end{itemize}

Here we would like to point out that naturally a cosmological horizon takes place in de Sitter spacetime.
In the vacuum conditions the cosmological horizon appears
at the first zero of $N(x)$. This occurs at $x_0^v  = \frac{\pi}{\sqrt{3\bar{\Lambda}}}$. At the same position,
$L(x\rightarrow x_0^v) \rightarrow \infty$.  We also want to mention that the singular behavior of the
functions $N(x)$ and $L(x)$ near $x_0^v$ is similar to the singular behavior of the corresponding components of
the metric tensor associated with supermassive configuration analyzed in \cite{Pablo}. Near their
corresponding singular point these functions behave as:
\begin{equation}
N(x)\approx(x_{sing}-x)^{2/3} \ {\rm and} \ L(x)\approx (x_{sing}-x)^{-1/3} \  .
\end{equation}
However we would like to emphasize that the physical reasons for both singular behaviors 
are different. The source of the singular behavior found in \cite{Pablo} is a supermassive
configuration of matter fields. Here is the presence of a positive cosmological constant. 
As to the anti-de Sitter spacetime there is no cosmological horizon.

Although the above analysis present important information about the behaviors
of the metric fields $N$ and $L$, we expect that the non-trivial structure of the Higgs
and gauge fields produce relevant modifications on these behaviors.\footnote{Specifically 
in the Minkowski spacetime, it is well known that the Abelian and also non-Abelian 
strings produce significant modifications in the geometry when compared with vacuum, the most relevant one
is associated with the decreasing in the slope of $L$ causing a planar angle
deficit.} We leave to the next section this analysis.

%
\section{Numerical Solutions}
\label{Numerical}
%
In this section we  shall analyze numerically  our system. To do that we integrate numerically
the equations (\ref{eq1}) - (\ref{eq5})  with the appropriated  boundary conditions specified in (\ref{eqbound1})-(\ref{eqbound5})
corresponding to dS and AdS cases,
by using the ODE solver COLSYS \cite{Colsys}. Relative errors of the functions
are typically on the order of $10^{-8}$ to $10^{-10}$ (and sometimes even better).

Our objective is to  analyze the behavior of the solutions of the non-Abelian cosmic string in de
Sitter and anti-de Sitter spacetime. In order to do this, we construct solutions by  specifying  the set of physical
parameters of the system for positives (de Sitter spacetime) and negatives (anti-de Sitter spacetime) values of cosmological
constant, $\bar\Lambda$. Moreover, we are also interested in comparing these behaviors with the corresponding one for the Abelian
gravitating strings, observing, separately, the influence of each system on the geometry of the spacetime.

 \subsection{ de Sitter Spacetime}

\begin{figure}[!htb]
\begin{center}
\includegraphics[width=0.9\textwidth]{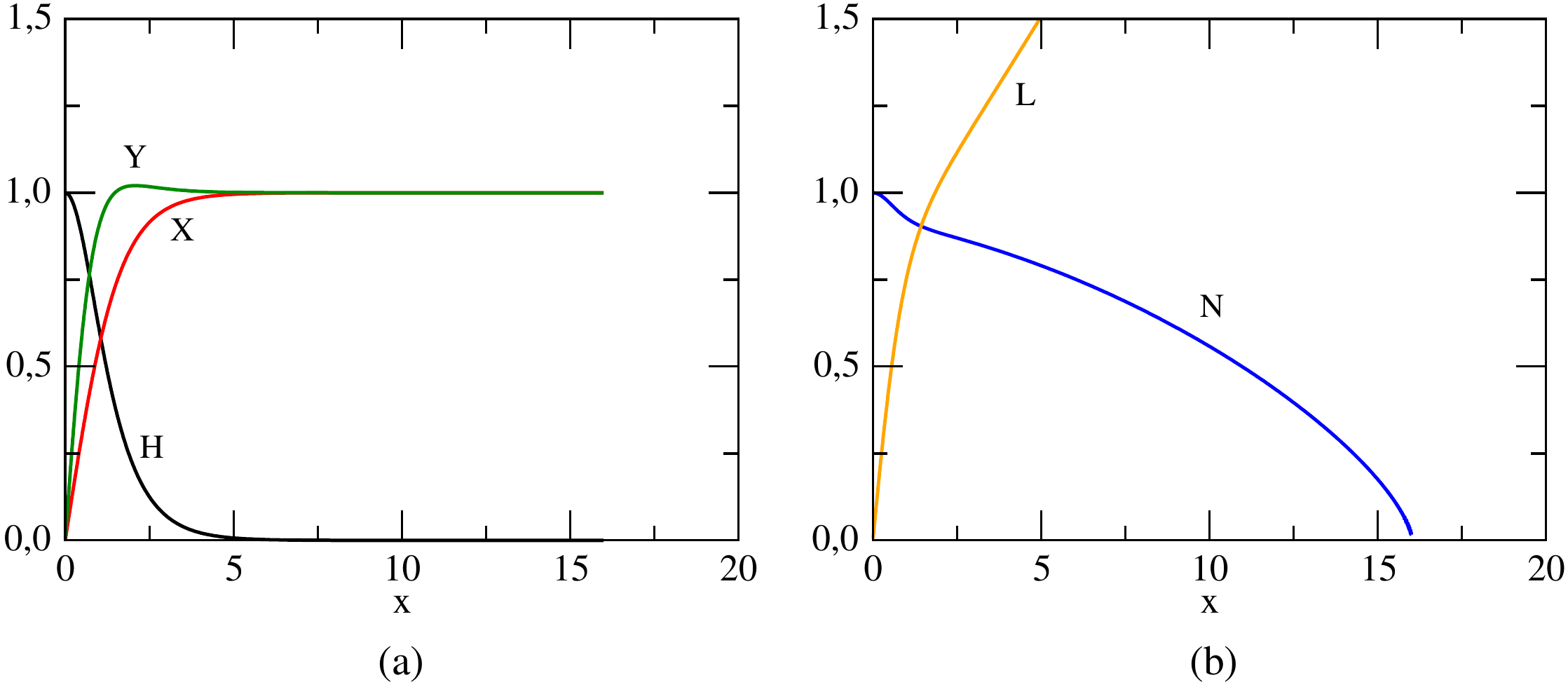}
\vspace{0.7cm}
\caption{\label{non_abelian} Non-Abelian string: In the left plot, we present the behavior for the Higgs and gauge fields
as functions of $x$. In the right plot, we present the behavior of the metric functions as functions of $x$. In both plots we consider
the parameters $\alpha=0.8$, $\gamma=0.61$, $\bar\Lambda = 0.0075,$ $\beta_2= 2.0$, $\beta_3=1.0$ and $q=1.0$.}
\label{fig1}
\end{center}
\end{figure}

In the first moment, we shall analyze the behaviors of Higgs, gauge and metric fields for the non-Abelian cosmic strings in de Sitter spacetime.
Our results for non-Abelian case are shown in figure \ref{fig1}. In the left plot we present the Higgs fields, $X$ and $Y$, and gauge
field, $H$,  as functions of $x$. In the right plot we present the behavior of metric functions, $N$ and $L$.  In both plots we set
the parameters as $\alpha=0.8$, $\gamma=0.61$, $\bar\Lambda = 0.0075,$ $\beta_2= 2.0$, $\beta_3=1.0$ and $q=1.0$.

In figure 2 we present the behavior of the Higgs and gauge fields, and the  metric functions for the Abelian case in de Sitter spacetime. In the
left plot, we exhibit the Higgs and gauge fields,  $X$ and  $H$, respectively, as function of  dimensioles variable  $x$.
In the right  plot we present the metric functions, $N$ and
$L$ as functions of $x$. For  both plots we consider the parameters $\alpha=0.8$, $\gamma=0.61$ and $\bar\Lambda = 0.0075$.

\begin{figure}[!htb]
\begin{center}
\includegraphics[width=0.9\textwidth]{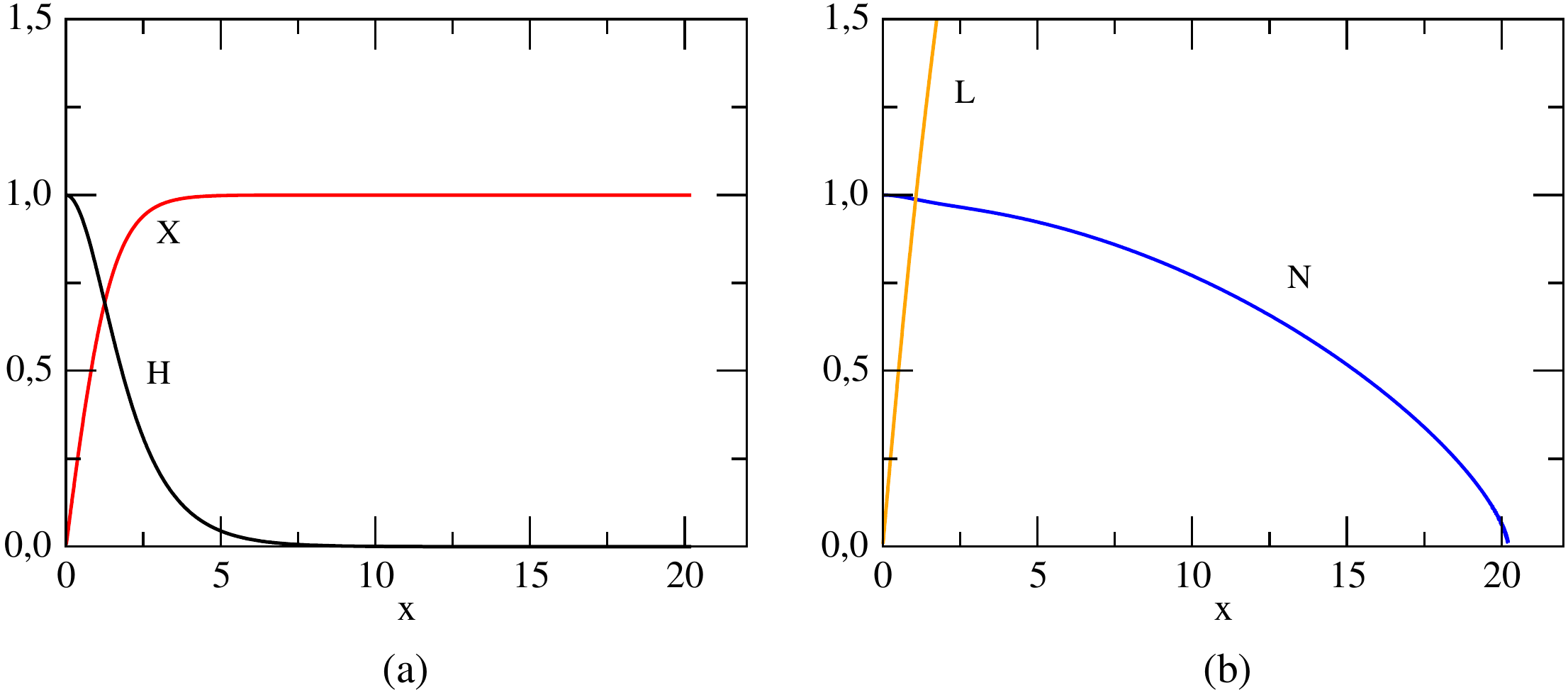}
\vspace{0.7cm}
\caption{\label{abelian}  Abelian string: In the left plot, we present the behavior for the Higgs and gauge fields as functions of $x$.
In the right plot, we present the behavior of the metric functions as functions of $x$. In both plots we consider the parameters
$\alpha=0.8$, $\gamma=0.61$, $\bar\Lambda = 0.0075$.}
\label{fig2}
\end{center}
\end{figure}
\newpage
By comparing figure 1(b) with figure  2(b) it can be seen that both systems present cosmological horizons, $x_0$. Moreover,
the corresponding values for the horizons for the  non-Abelian system is smaller than the  Abelian
one. Another point that can be mentioned is that, for  values of the parameters that we have chosen,
 the matter fields and gauge fields  reach their asymptotic values in the region inside the horizons.

In the vacuum solution we have found that the cosmological horizon is reached for $x_0^v=\frac\pi{\sqrt{3\bar{\Lambda}}}$,
which for the value of constant cosmological adopted in the plots, provides  $x_0^v\approx20.94395$.
More realistic values for the cosmological horizons were obtained in both  plots considering the non-trivial behaviors of the fields.
Motived by this fact now we want to investigate how
the cosmological horizon depends on the gravitational coupling,   $\gamma$, and also on the cosmological constant itself, $\bar{\Lambda}$.

 First we consider the dependence of $x_0$ with $\gamma$.
  In order to make this analysis we fixed $\alpha$,
$\beta_2$, $\beta_3$, $\bar\Lambda$ and $q$. The value of the
cosmological horizon is obtained when the metric function $N(x=x_0)$ is zero. Our numerical results for $\alpha = 0.8,
\beta_2 = 2.0, \  \beta_3 = 1.0, \ \bar\Lambda = 0.0075$ and $q = 1.0$ are presented in figure 3(a). Note that the cosmological horizon decreases
as the value of $\gamma$ is increased.

As to the influence of $\bar\Lambda$ on the cosmological horizon, we adopted a similar
procedure to the case above. Nevertheless we fixed $\alpha$, $\gamma$, $\beta_2$, $\beta_3$ and $q$ and we determine the
value of $x$ at which $N(x)$ vanishes. Our results are presented in figure 3(b). We clearly note that the value of cosmological
constant decreases as values of $\bar\Lambda$ is increased. Also in this plot, we provide the behaviour for the
cosmological horizon in the vacuum case.

\begin{figure}[!htb]
\begin{center}
\includegraphics[width=0.9\textwidth]{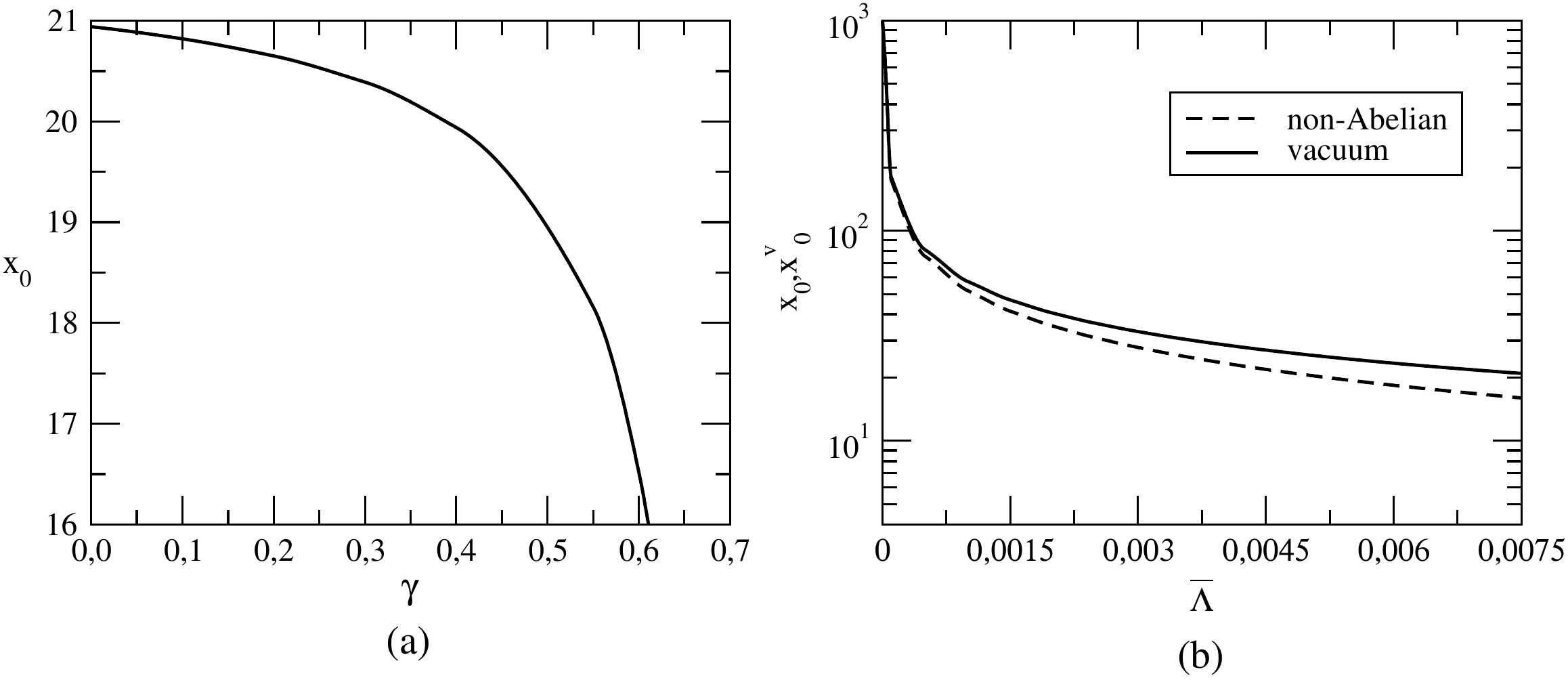}
\vspace{0.7cm}
\caption{(a) The behavior of the cosmological horizon, $x_0$, as
function of $\gamma$ considering the parameters $\alpha = 0.8, \beta_2 = 2.0, \beta_3 = 1.0, \bar\Lambda = 0.0075, q = 1.0$.
(b) The dashed line represents the behavior of the cosmological horizon, $x_0$, as a function of $\bar\Lambda$ for
the non-Abelian string case, considering $\alpha = 0.8, \gamma = 0.61, \beta_2 = 2.0, \beta_3 = 1.0, q = 1.0$.
The solid line corresponds the trivial behavior of the cosmological horizon, $x_o^v$, in the vacuum case. }
\label{fig3}
\end{center}
\end{figure}

 \subsection{ Anti-de Sitter Spacetime}

Here, we are interested to analyze the influence of a negative cosmological
constant on the behavior of the  non-Abelian and Abelian cosmic string systems.

In figure 4(a) we present the behavior of the Higgs fields,
$X, \  Y$, and gauge field, $H$, for the non-Abelian case as function of the dimensionless variable $x$,
considering specific values attibuted to the set of parameters. In figure 4(b) we plot the behavior of the
corresponding metric functions, $N$ and $ L$
as function of $x$. In both plots we considered the parameters
$\alpha = 0.8,  \ \gamma = 0.6,  \  \beta_2 = 2.0,  \  \beta_3 = 1.0,  \  \bar\Lambda = -0.03$ and $  q = 1.0$.

In figure 5(a) we present the behavior of the Higgs field, $X,$ and gauge field, $H,$ in the Abelian case for specific values
attributed to set of  appropriated parameters. In figure 5(b) we plot the metric functions, $N$ and $L$.
In both plots we consider the parameters
for Abelian case as $\alpha = 0.8, \  \gamma = 0.6$ and $\bar\Lambda = -0.03$.
For both systems we can see that the function $N$ presents a strong increment for large value of $x$.

Finally we present in figure 6  the behavior of the metric fields for two different values of cosmological constant,
$\bar\Lambda=-0.010$ and $\bar\Lambda=-0.007$, in the non-Abelain case with parameters
$\alpha = 0.8, \gamma = 0.6, \beta_2 = 2.0, \beta_3 = 1.0$ and $q = 1.0$. We notice
that the metric field $N$ increases with the cosmological constant.

\begin{figure}[!htb]
\begin{center}
\includegraphics[width=0.9\textwidth]{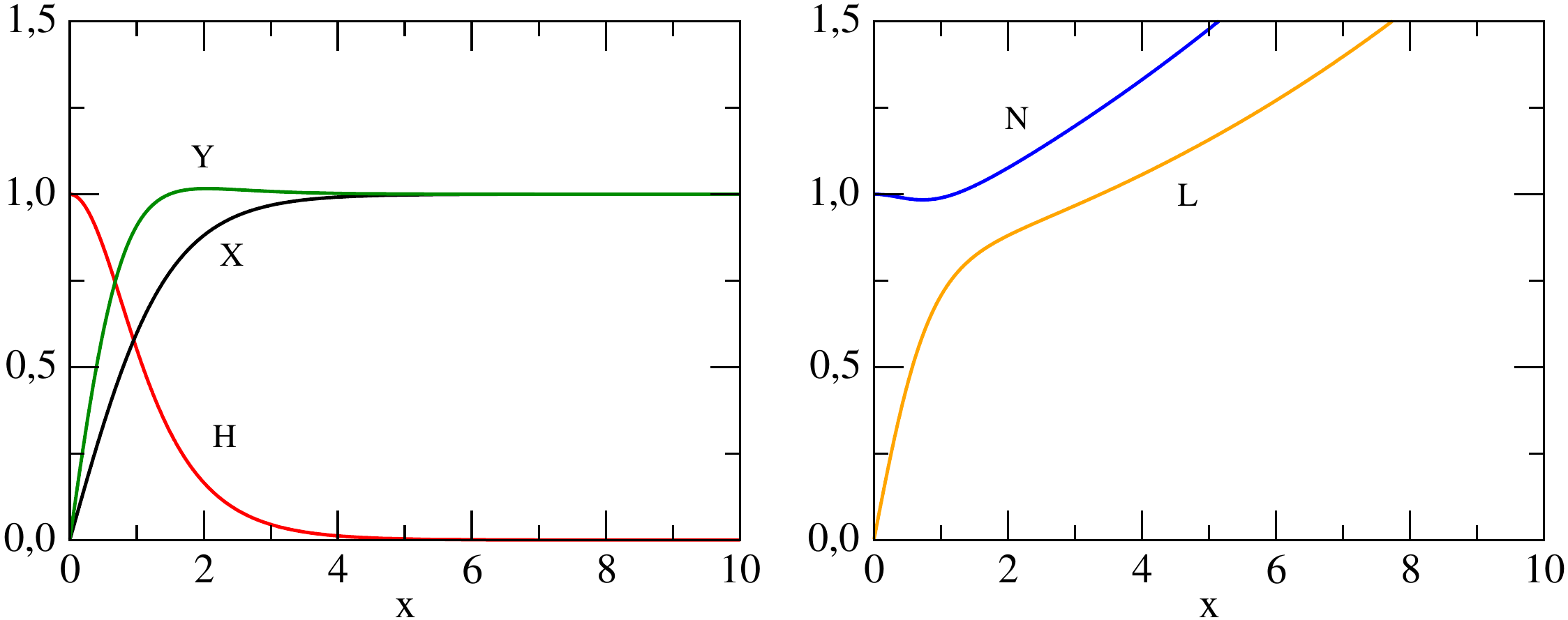}
\vspace{0.7cm}
\caption{Non-Abelian string: In the left  plot we exhibit the behavior of the Higgs and gauge fields
in anti-de Sitter space. In the right plot, we present the metric functions. In both plots we have considered
$\alpha = 0.8, \  \gamma = 0.6,  \  \beta_2 = 2.0, \   \beta_3 = 1.0,
\bar\Lambda = -0.03$ and $ q = 1.0$.}
\label{fig4}
\end{center}
\end{figure}
\begin{figure}[!htb]
\begin{center}
\includegraphics[width=0.9\textwidth]{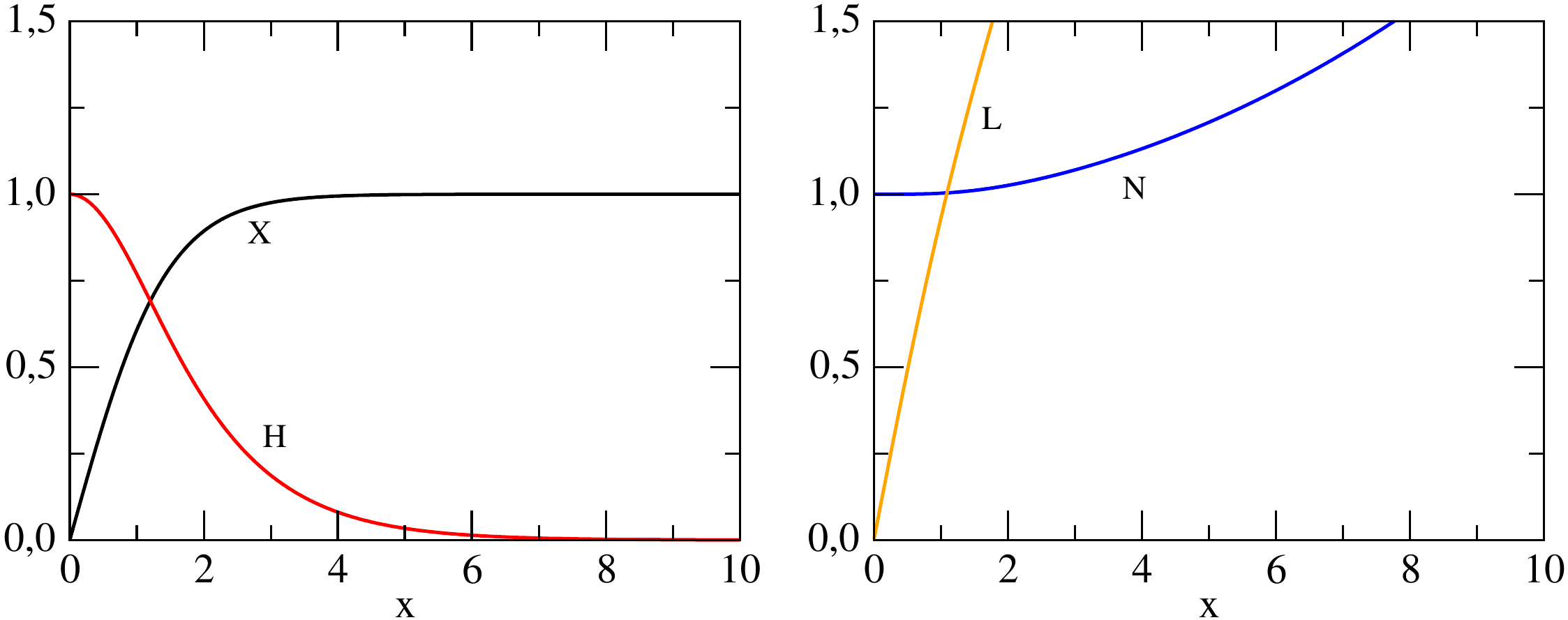}
\vspace{0.7cm}
\caption{Abelian string: In the left plot we exhibit the behavior of the  Higgs and gauge fields in anti-de Sitter space
as function of $x$. In the right plot, we exhibit the behavior of the  metric fields
 in anti-de Sitter space considering $\alpha = 0.8, \gamma = 0.6, \bar\Lambda = -0.03$.}
\label{fig5}
\end{center}
\end{figure}
\begin{figure}[!htb]
\begin{center}
\includegraphics[width=0.6\textwidth]{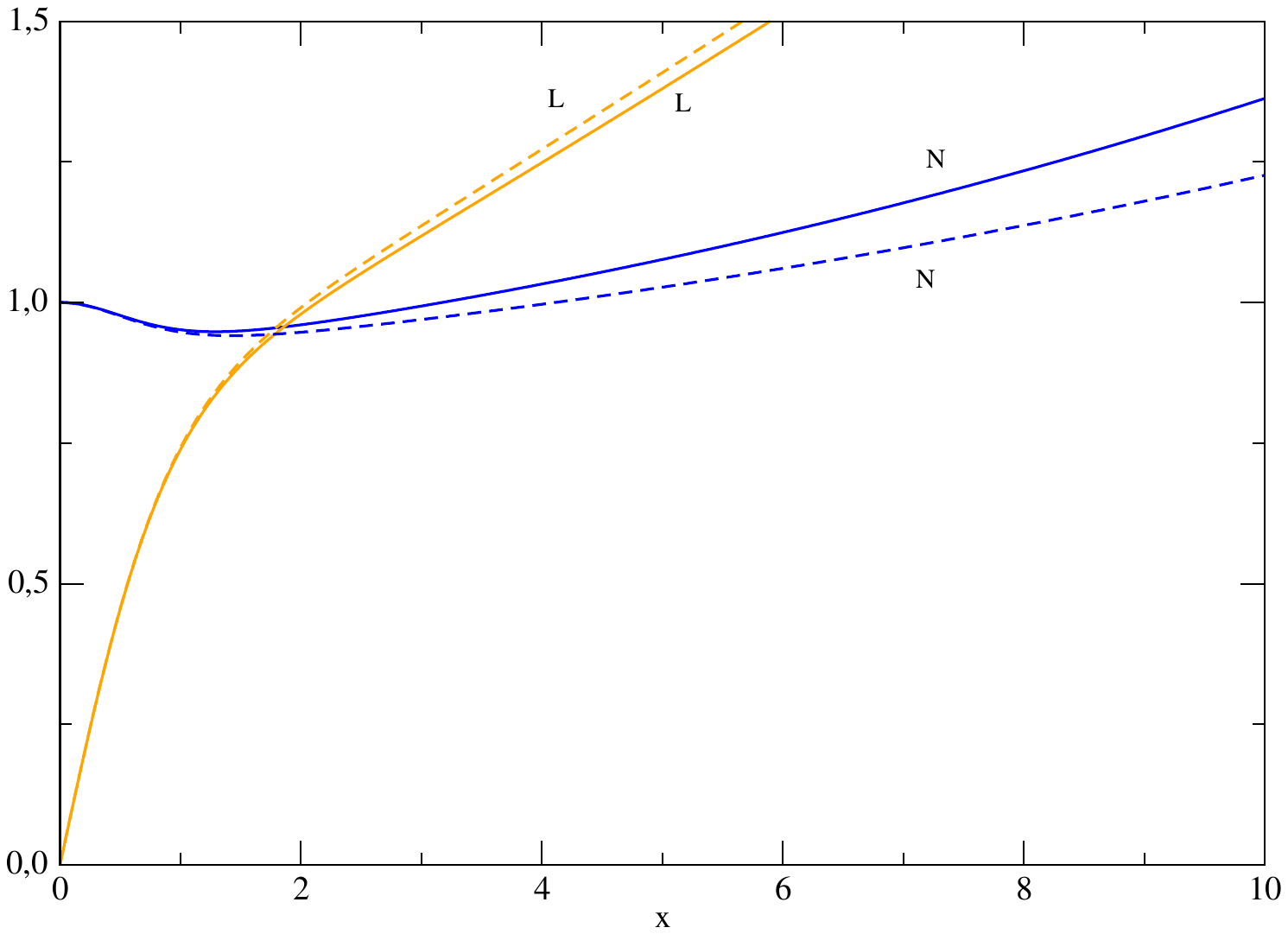}
\caption{The metric fields as functions of $x$ for different values of cosmological constant. 
For solid lines we adopt $\bar\Lambda=-0.010$, and for
dashed lines $\bar\Lambda=-0.007$. In both plots we consider $\alpha = 0.8, \gamma = 0.6, \beta_2 = 2.0, \beta_3 = 1.0$ and  $q = 1.0$.}
\label{fig6}
\end{center}
\end{figure}

 \subsection{Comparative analysis}
\label{Comp}
In this section we would like to present plots comparing the behaviors of the metric fields, 
$L(x)$ and $N(x)$, as functions of $x$ for different background spacetimes: $a)$ the non-Abelian cosmic sting
in Minkowski ({\it M}) and in de Sitter ({\it dS}) spacetimes, and $b)$ the non-Abelian cosmic string
in Minkowski ({\it M}) and anti-de Sitter ({\it AdS}) spacetimes. In addition 
we have included the behaviors of these metric functions in the vacuum ({\it vac}) for {\it dS} and {\it AdS}
spacetimes, given by equations \eqref{NdS}, \eqref{LdS}, \eqref{NAdS} and \eqref{LAdS}. In the plots we
label these curves by the subscripts $M$, $dS$, $AdS$ and $vac$, according to the space background considered. 
In this way we intend to point out the most relevant aspects that distinguish the behaviors of those
functions. In our analysis presented in figure 7, we adopted the following values for the parameters:
 $\gamma=0.6, \ \alpha = 0.8, \ \beta_2 = 2.0, \ \beta_3 = 1.0$ and  $q = 1.0$. For dS space we take $\bar\Lambda=0.0075$
and for AdS we take $\bar\Lambda=-0.03$. The behaviors of the
Higgs and gauge fields are almost insensitive to the presence of a cosmological constant, for 
this reason we decided do not include them in the plots. 

In figure 7(a) we present the behaviors of $L$ and $N$ as function of $x$ considering the non-Abelian
cosmic string in Minkowski and in de Sitter spectimes. Also we present their behaviors in vacuum of de Sitter space, named
vacuum solutions. 
We can see that the main difference is in the behaviors of the component $N$. In Minkowski space this component tends to a constant 
value below the unity while in dS space it goes to zero. In addition, comparing
$N$ in dS with in the vacuum, we can see that the cosmological horizon for the first case
is smaller that the second one. A less evident difference is in the behavior of $L$. Comparing the
plot of this component in the vacuum, $L_{vac}$, in the non-Abelian string in dS, $L_{dS}$, and 
non-Abelian string in Mikowski, $L_M$,
spacetimes respectively, we can notice a progressive bending. Specifically there is a small deviation between
$L_{dS}$ and $L_{M}$. Another point that deserves to be mentioned is the decreasing in the slope of $L$ when one compares
$L_{vac}$ with $L_{dS}$. The slope of the latter is smaller for any given point. This resembles the decreasing in the slope of $L_M$ when compared
with the one in the vacuum in Mikowski spacetime. In fact comparing the slope of $L_M$ at infinity with the unity, 
it is possible to find a planar angle deficit by:
\begin{equation}
\delta/2\pi=1-L'_M(\infty)=0.865 \  .
\end{equation}  

In figure 7(b) we present the behaviors of $L$ and $N$ as function of $x$ considering the non-Abelian
cosmic string in Minkowski and in anti-de Sitter spectimes. In addition we present their behaviors
in vacuum of anti-de Sitter space. Here also we can see that the main difference in the geometry of the spacetime
is given by $N$: $N_{AdS}$ increases with $x$ while $N_M$ presents a small decay. 
As to $L$ we observe that for a given value of $x$, $L_M$ is bigger than $L_{AdS}$; moreover, 
both are smaller that $L_{vac}$. So, we conclude that the non-trivial structure of the Higgs and gauge fields
modify the behavior of $L$. Specifically the slope of $L_{AdS}$ is smaller that $L_{vac}$.

So, from these two plots, figures 7(a) and 7(b), two different observations deserve to be mentioned:
\begin{itemize}
\item The presence of a cosmological constant affects substantially the geometry of the non-Abelian cosmic string spacetime  modifying mainly the components $g_{tt}=g_{zz}$ of the metric tensor. 
\item In the other direction we can see that the presence of matter field in dS and AdS spaces also produce relevant consequences on the behavior of the metric tensor. Specifically in dS space the value of the cosmological horizon decreases significantly.   
\end{itemize}

\begin{figure}[!htb]
\begin{center}
\includegraphics[width=0.9\textwidth]{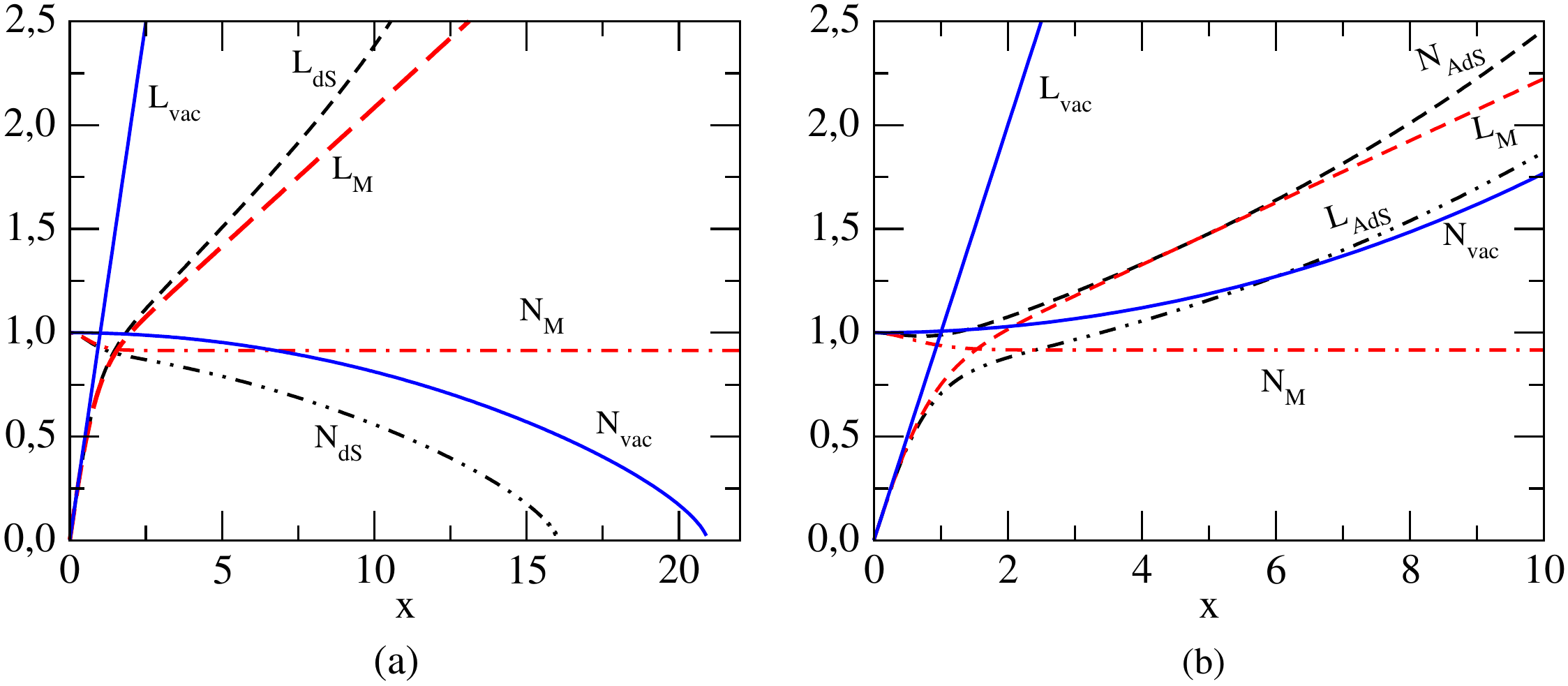}
\caption{The metric fields $L(x)$ and $N(x)$ as functions of $x$ . (a) In figure 7(a) we compare the behavior 
of the metric fields $L$ and $N$ in de Sitter spacetime ({\it dS}) considering  $\bar\Lambda=0.0075$ 
with the metric fields in the Minkowski ({\it M}),
and with the metric fields in the respective vacuum configuration ({\it vac}).
(b)  In figure 7(b) we compare the behaviour of the metric fields $L$ and $N$ in anti-de Sitter spacetime ({\it AdS}) 
considering  $\bar\Lambda=-0.03$ with the metric fields in the Minkowski ({\it M})
and with the metric fields in the respective vacuum configuration ({\it vac}). In both plots we consider
 $\gamma=0.6, \ \alpha = 0.8, \beta_2 = 2.0, \beta_3 = 1.0$ and  $q = 1.0$.}
\label{fig6}
\end{center}
\end{figure}

\section{Conclusion}
\label{Concl}
%
In this paper, we have examined the influence of cosmological constant in the 
geometry of non-Abelian and Abelian cosmic strings spacetimes. In agreement
with previous work \cite{Ant_Eugenio},  where the gravitating non-Abelian cosmic strings was studied in the absence of cosmological
constant, we have shown that it is also be possible to obtain non-Abelian stable topological string considering two bosonic
iso-vectors with Higgs mechanism in de Sitter and anti-de Sitter spaces.

Regarding the analysis in de Sitter space, we have shown that there appear a cosmological horizon. In fact this
observation  was presented in the paper by Linet for the vacuum configuration in \cite{Linet}, and in \cite{eugenio2}
for the Abelian string. Here we also returned to this analysis  considering the non-Abelian
and also Abelian strings. These investigations were presented in figures 1(b) and 2(b) for specific values of the parameters.
By these graphs we have pointed out that the
 non-Abelian case presents a smaller cosmological horizon than corresponding Abelian one.

We also provided the behaviour of cosmological horizon, $x_0$, with the gravitational 
coupling constant, $\gamma$, and with the cosmological
constant, $\bar\Lambda$. In figure 3(a), we can observe that the cosmological horizon decreases when one increases $\gamma$.
In figure 3(b) we can see that  the cosmological
horizon also decreases for larger values of the cosmological constant.
Moreover, we also compare this behavior with the corresponding one for the vacuum case. We see
that for a given value of $\bar\Lambda$, the horizon associated with the vortex system is smaller than the vacuum one.

The behaviors of the Higgs and gauge fields, and metric functions in anti-de Sitter spacetime for the non-Abelian cosmic strings
were displayed in figure 4. We have also shown the behaviors of Higgs and gauge fields, 
and metric functions for the Abelian cosmic strings in figure 5. We noticed that
in both cases $N(x)$ diverges for large values of $x$; however, by our numerical results, we 
observe that in the non-Abelian case the slope of $N(x)$ is
bigger than the corresponding  Abelian one. In the figure 6, we have 
shown the behavior of metric fields, $N$ and $L$, as functions of $x$ for the non-Abelian
system considering two different values of $\bar\Lambda$.  
By this graph and others not presented in this paper,  we
observe that increasing the cosmological constant the two lines, representing these functions, 
approach each other. This behavior is compatible  with the vacuum case.

Finally we have presented in figures 7(a) and 7(b) comparative plots of the behaviors of the components $N(x)$
and $L(x)$ of the metric tensor as function of $x$, considering the non-Abelian string in Minkowski and in
de Sitter backgrounds, and in Mikowski and in anti-de Sitter backgrounds, respectively. We have observed that the presence of cosmological constant
strongly modifies the geometry of the spacetimes produced by the  defect. The most relevant
modifications are due to the component $N(x)$. For dS, $N$ goes to zero for a finite distance to
the string, and for AdS this component increases. Also in these graphs we have presented the behaviors
of these two functions in vacuum scenarios. By comparison of the functions in the vacuum
scenarios with the full system in dS or AdS, we have observed significant deviations 
caused by the Higgs and gauge fields in the slopes of $L$.  
%
\section*{Acknowledgments}
ERBM would like to acknowledge CNPq for partial financial support. AdPS would like to acknowledge the Universidade
Federal Rural de Pernambuco.
%


\end{document}